\documentclass[a4paper,twocolumn,english,prl,superscriptaddress,showpacs,tight]{revtex4}
\usepackage{ae,aecompl}
\usepackage[T1]{fontenc}
\usepackage[latin1]{inputenc}
\usepackage{float}
\usepackage{textcomp}
\usepackage{amsmath}
\usepackage{amssymb}
\usepackage{graphicx}
\usepackage{esint}

\makeatletter

\pdfpageheight\paperheight
\pdfpagewidth\paperwidth

\@ifundefined{textcolor}{}
{%
 \definecolor{BLACK}{gray}{0}
 \definecolor{WHITE}{gray}{1}
 \definecolor{RED}{rgb}{1,0,0}
 \definecolor{GREEN}{rgb}{0,1,0}
 \definecolor{BLUE}{rgb}{0,0,1}
 \definecolor{CYAN}{cmyk}{1,0,0,0}
 \definecolor{MAGENTA}{cmyk}{0,1,0,0}
 \definecolor{YELLOW}{cmyk}{0,0,1,0}
}

\usepackage{epsfig}
\usepackage{bm}
\usepackage{babel}
\usepackage{amsmath}
\usepackage{amssymb}
\usepackage{graphicx}
\usepackage{esint}

\pdfoutput = 1

\makeatother

\usepackage{babel}
\begin{document}

\title{Complete light absorption in graphene-metamaterial corrugated structures}

\author{Aires Ferreira}

\affiliation{Graphene Research Centre and Department of Physics, National University
of Singapore, 2 Science Drive 3, Singapore 117542}

\author{N. M. R. Peres}

\affiliation{Department of Physics and Center of Physics, University of Minho,
P-4710-057, Braga, Portugal}
\begin{abstract}
We show that surface-plasmon polaritons excited in negative permittivity
metamaterials having shallow periodic surface corrugation profiles
can be explored to push the absorption of single and continuous sheets
of graphene up to 100\%. In the relaxation regime, the position of
the plasmonic resonances of the hybrid system is determined by the
plasma frequency of the metamaterial, allowing the frequency range
for enhanced absorption to be set without the need of engineering
graphene. 
\end{abstract}


\maketitle

\section{Introduction\label{sec:Introduction}}

The presence of a boundary in a metal allows for surface collective
charge density oscillations in addition to volume plasmon modes.\cite{Ritchie1957}
When these surface modes couple to light, hybrid photon-plasma excitations
emerge, known as surface-plasmon polaritons (SPPs).\cite{Economou1969,Agranovich1982,Burke1986,Hetch1996} 

A notable feature of a SPP is its small wavelength, when gauged against
the wavelength of a transverse photon with the same frequency. This
results in high degrees of electromagnetic energy confinement, endowing
SPPs with high sensitiveness to surface conditions.\cite{Agranovich1982,Zayatsa2005}
For this reason, SPPs are responsible for surface-enhanced Raman scattering,\cite{Moskovits1985}
enabling single molecule detection,\cite{Kneipp1997} and having many
technological applications (e.g.,~chemical and biological sensors,\cite{Homola1999}
and nano-resolution imaging\cite{Gramotnev2010}). Moreover, the compact
nature of SPPs offers the prospect of combining large bandwidth and
integration, at subwavelength scale, eventually allowing nanoscale
photonic circuit implementation.\cite{Barnes2003,Gramotnev2010} 

The scope of surface plasmon-related phenomena has been increasing
swiftly in the past decade, spanning metallic nanowires,\cite{Weeber1999}
bandgap nanostructures,\cite{Bozhevolnyi2001} metallic nanoparticles,\cite{Kelly2003}
and other nano-engineered materials.\cite{Henzie2007,Williams2008,Kabashin2009}
With the advent of \emph{truly} two-dimensional crystals---graphene
being the first of the kind\cite{Novoselov2005,Novoselov2012}---a
new playground for plasmonics has emerged.\cite{Koppens} The unique
electronic properties of graphene, characterized by massless and chiral
low-energy excitations,\cite{rmp} and unconventional transport properties,\cite{colloquium}
originate the most distinct electromagnetic confinement behavior:
SPPs with propagation lengths exceeding those of conventional metal-dielectric
interfaces,\cite{plasmonics_graph_zerofield_Jablan} and guided transverse-electric
modes,\cite{ZieglerPRL,MPPQTE} just to mention a few.

The observation of prominent plasmonic absorption peaks in graphene
microarrays\cite{Ju_MicroArrays} has triggered a new research line,\cite{Nikitin2012,Koppens2012}
in which plasmonic excitations are explored to overcome the major
obstacle in graphene-based optoelectronics: the small light absorption
in one-atom thick graphene samples.\cite{Nair2008} Inspired by these
works and well-established results for metallic gratings,\cite{Zayatsa2005}
we propose a hybrid graphene-metamaterial system, where a single and
continuous graphene sheet is seen to absorb all the light impinging
on it. The frequency range for enhanced absorption is determined by
the metamaterial alone (see later), which allows subsequent control
over the absorbed plasmonic waves to be performed in the graphene
sheet (e.g.,~via chemical doping\cite{Avouris2012}).

\begin{figure}[H]
\centering{}\includegraphics[clip,width=0.8\columnwidth]{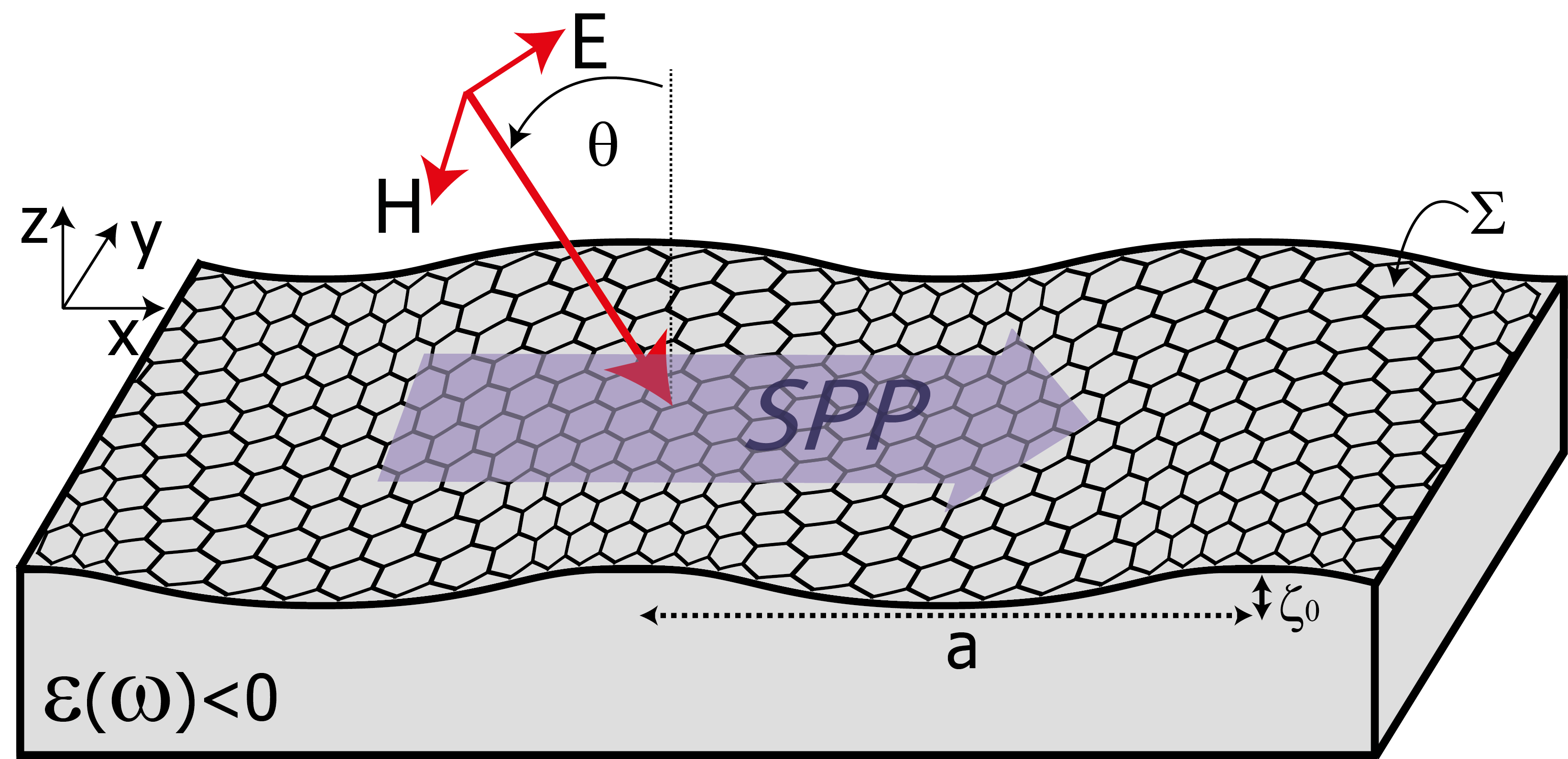}\caption{\label{fig:01}Schematic picture showing transverse magnetic light
impinging on graphene placed on a periodically structured conducting
substrate. The light is converted into surface plasmonic waves and
subsequently absorbed in graphene. }
\end{figure}

The system we have in mind is depicted in Fig.~\ref{fig:01}: incident
light with angular frequency $\omega$ strikes the interface $\Sigma$
of a periodically structured conducting system coated with graphene.
The light is transverse magnetic polarized, $\boldsymbol{H}_{\textrm{in}}(\boldsymbol{r},t)=H_{\textrm{in}}e^{i(\boldsymbol{k}\cdot\boldsymbol{r}-\omega t)}\boldsymbol{e}_{y}$
, where $\boldsymbol{k}=(k_{x},0,k_{z})$, and the parametrization
of the boundary reads $\Sigma=\{(x,y,z)|z=\zeta(x)\}$, with $\zeta(x)$
having the symmetry property, $\zeta(x+a)=\zeta(x)$. In order to
achieve electromagnetic energy confinement and SPP-related phenomena
at the interface $\Sigma$, the substrate is assumed to have $\epsilon(\omega)<0$
in the frequency domain of interest.\cite{Zayatsa2005} In a practical
implementation, the latter would be an engineered metamaterial,\cite{Pendry1996}
in the relaxation regime $\omega_{p}\gg\omega\gg\gamma$ (with $\omega_{p}$
being the effective plasma frequency and $\gamma$ the relaxation
rate). 

The spatial periodicity of the present configuration is instrumental
in order to excite SPPs via coupling to the radiation field: conservation
of momentum requires that the projection of the wavevector of the
incident photons onto the plane parallel to $\Sigma$ matches the
wavevector $q$ of SPPs belonging to the hybrid structure. In flat
surfaces the matching condition cannot be achieved since, in that
case, the SPP dispersion relation lies to the right of the light line
(i.e.,~$q>\omega/c\ge k_{x}$). The role of the periodic corrugated
profile is thus to open Bragg scattering channels providing photons
with the necessary wave vector (according to $q=k_{x}\pm2\pi n/a$,
for some integer $n$), thus allowing the excitation of SPPs to occur
via direct illumination.

\section{Light Scattering: Exact Integral-Equation Formulation\label{sec:Formalism}}

The investigation of the optical properties of the hybrid graphene-metamaterial
structure (Fig.~\ref{fig:01}) is carried out via exact numerical
calculations of the reflected and transmitted fields. The total magnetic
field is given by $H_{y}(\boldsymbol{r},t)=H_{\pm}(\boldsymbol{r})e^{-i\omega t}$
{[}here $\boldsymbol{r}=(x,z)$, and $\pm=\textrm{sign}\,(z-\zeta(x))${]},
where the amplitudes $H_{\pm}(\boldsymbol{r})$ are solutions of the
wave equation,$\left[\nabla^{2}+\epsilon_{\pm}(\omega)\omega\text{\texttwosuperior}/c^{2}\right]H_{\pm}(\boldsymbol{r})=0$,
with $\nabla^{2}=\partial_{x}^{2}+\partial_{z}^{2}$, and where $\epsilon_{+}(\omega)=1$
(vacuum) and $\epsilon_{-}(\omega)=1-\omega_{p}^{2}/[\omega(\omega+i\gamma)]$
(substrate). The boundary conditions at the interface $\Sigma$ read
as\cite{comment_boundary_condition}
\begin{eqnarray}
H_{-}(x,\zeta^{-})-J_{\Sigma} & = & H_{+}(x,\zeta^{+})\equiv H(x)\,,\label{eq:boundary_cond_1}\\
\frac{1}{\epsilon_{-}}\partial_{n}H_{-}(x,\zeta^{-}) & = & \frac{1}{\epsilon_{+}}\partial_{n}H_{+}(x,\zeta^{+})\equiv\tilde{L}(x)\,,\label{eq:boundary_cond_2}
\end{eqnarray}
where the notation $\zeta^{\pm}$ means that the fields (and derivatives)
are to be evaluated taking values of $z$ approaching $\Sigma$ from
above ($\zeta^{+}$) or below ($\zeta^{-}$) , $\partial_{n}\equiv\boldsymbol{n}\cdot\boldsymbol{\nabla}$
denotes the derivative along the unit vector normal to the surface
$\boldsymbol{n}$, directed from the vacuum into the substrate, and
$J_{\Sigma}$ is the longitudinal current induced by graphene. 

The surface current relates to the electromagnetic fields according
to the constitutive relation (Ohm's law), $J_{\Sigma}\equiv J(x)=\sigma(x,\omega)\boldsymbol{E}(x,\zeta)\cdot\boldsymbol{t}$,
where $\sigma(x,\omega)$ is the optical conductivity of the corrugated
graphene sheet, which in general may depend on the $x$ coordinate,
and $\boldsymbol{E}(x,\zeta)\cdot\boldsymbol{t}=i\tilde{L}(x)/(\varepsilon_{0}\omega)$
is the tangential component of the electric field evaluated at $\boldsymbol{r}=(x,\zeta(x))$.
Here, $\boldsymbol{t}$ denotes the unit vector tangent to $\Sigma$
and $\varepsilon_{0}$ is the vacuum permittivity. In what follows,
we approximate the optical conductivity by its bare value $\sigma(\omega)$
in the absence of corrugation. This is justified since the inclusion
of a spatial modulation in $\sigma(\omega)$ does not significantly
alter the optical properties to be discussed throughout (see Sec.~\ref{sec:Final_Remarks}).
We take $\sigma(\omega)$ as obtained by the random-phase approximation,\cite{colloquium}
with a relaxation rate $\Gamma=2.5$~meV, in consistency with typical
values.\cite{pump_probe_epitaxial,pump_probe_exfoliated,Horgn2011} 

In order to solve for the full electromagnetic field we make use of
an exact integral-equation method developed by Toigo \emph{et al}.\cite{Toigo1977}
This method exploits the integral form of Maxwell equations in order
to express $H_{\pm}(\boldsymbol{r})$ as an integral over the functions
$H(x)$ and $L(x)\equiv\tilde{L}(x)\sqrt{1+[\zeta^{\prime}(x)]{}^{2}}$.
Making use of the Bloch property, $H(L)=\sum_{n}H(L)_{n}e^{iq_{n}x}$,
with $q_{n}\equiv k_{x}+2\pi n/a$, Toigo\emph{ }and co-workers reduce
the well-known integral equations\cite{BornAndWolf} to a set of linear
algebraic equations for the Fourier coefficients of $H_{\pm}(\boldsymbol{r})$.
Here, we generalize their method as to include the effect of a metallic
sheet with conductivity $\sigma(\omega)$ deposited on the corrugated
surface $\Sigma$. We just state the basic results: above (below)
the selvedge region $z>\max\zeta$ ($z<\min\zeta$), the reflected
($s=+$) {[}transmitted ($s=-$){]} wave is given by $\sum_{m=-\infty}^{\infty}\mathcal{H}_{m}^{s}e^{iq_{m}x}e^{-s\kappa_{s}(q_{m})z}$,
where the $m=0$ term corresponds to the specular wave, and the remaining
channels correspond to propagating modes (Bragg beams) or modes confined
to the surface $\Sigma$ (evanescent waves), depending on whether
\begin{equation}
\kappa_{s}(q_{m})=\sqrt{q_{m}^{2}-\epsilon_{s}(\omega)\frac{\omega\text{\texttwosuperior}}{c^{2}}}\quad,\textrm{with}\:\textrm{Im\,}\kappa_{s}\le0,\label{eq:decay}
\end{equation}
is pure imaginary or real, respectively. The amplitudes $\mathcal{H}_{m}^{s}$
relate to the Fourier coefficients $\{H_{n},L_{n}\}$ of the functions
defined in Eqs.~(\ref{eq:boundary_cond_1}) and (\ref{eq:boundary_cond_2})
according to
\begin{align}
\mathcal{H}_{m}^{s} & =\frac{1}{2\kappa_{s}(q_{m})}\sum_{n=-\infty}^{\infty}\Upsilon_{m}^{s-}(m-n)\left[\epsilon_{s}(\omega)L_{n}-\right.\nonumber \\
 & \left.s\frac{\epsilon_{s}(\omega)\omega^{2}/c^{2}-q_{n}q_{m}}{\kappa_{s}(q_{m})}\left(H_{n}+\delta_{s,-}\frac{i\sigma(\omega)}{\varepsilon_{0}\omega}L_{n}\right)\right],\label{eq:reflected_transmitted_fieldds}
\end{align}
where~$\Upsilon_{m}^{\alpha\beta}(p)\equiv a^{-1}\int_{-a/2}^{a/2}dxe^{-2\pi ipx/a}e^{-\alpha\beta\kappa_{\alpha}(q_{m})\zeta(x)}$
($\alpha,\beta=\pm1$) encodes the effect of the spatial profile of
the interface, and\begin{widetext}
\begin{eqnarray}
\sum_{n=-\infty}^{\infty}\Upsilon_{m}^{s+}(m-n)\left[L_{n}+s\frac{\epsilon_{s}(\omega)\omega^{2}/c^{2}-q_{n}q_{m}}{\epsilon_{s}(\omega)\kappa_{s}(q_{m})}\left(H_{n}+\delta_{s,-}\frac{i\sigma(\omega)}{\varepsilon_{0}\omega}L_{n}\right)\right] & = & 2ik_{z}H_{\textrm{in}}\delta_{m,0}\delta_{s,+}\,\label{eq:system_eq}
\end{eqnarray}
\end{widetext}provides an infinite set of linear equations for the
unknown coefficients $\{H_{n},L_{n}\}$.\cite{comment_approximation}
This system of equations is solved by retaining a finite number of
coefficients,~$n=-N+1\,...\, N$. The integer $N$ must be chosen
sufficiently large so that convergence is obtained. The actual convergence
properties of the present method depend crucially on the form of $\zeta(x)$,
and, in general, for corrugations with very large amplitudes and/or
possessing discontinuities, convergence is not guaranteed.\cite{Garcia1978,Agassi1986}
Here, we focus on periodic structures with a sinusoidal profile, $\zeta(x)=\zeta_{0}\cos\left(2\pi x/a\right)$,
for which the functions $\Upsilon_{m}^{\alpha\beta}(p)$ have a well-known
form and convergence is fast, i.e.,~only a few Fourier coefficients
are required for good accuracy.\cite{comment_convergence}

\section{Main results\label{sec:Results}}

We focus our discussion on normal incidence ($\theta=0$), with the
important remark that the reported features can be observed at $\theta>0$.
In that case, incident photons have $k_{x}=0$, and hence the wavevectors
of SPPs belonging to the hybrid structure coincide with the reciprocal
lattice wavevectors, $q_{n}=G_{n}\equiv2\pi/a$. In the relaxation
regime and in the ideal scenario of no losses occurring inside the
metamaterial (except when stated otherwise, we assume $\gamma=0$),
the fields cannot propagate into the substrate, and hence the incident
light is totally reflected at the interface (in the absence of graphene).
Yet, signatures of coupling of the incident light to SPPs are observed
in the zeroth order (specular) reflectivity of metallic gratings.\cite{Sheng1982,Weber_Mills_1983,Zayatsa2005}
The signatures arise because surface plasmonic waves excited at vacuum-metal
structured interfaces via direct illumination are \emph{radiative}:
the excited SPPs reradiate into the vacuum via open Bragg channels
($\omega>c|G_{n}|$, with $|n|>0$). The resulting general decrease
in the specular reflectivity is accompanied by pronounced features
(e.g.,~dips and bumps) appearing at the SPPs resonances.

\begin{figure}
\begin{centering}
\includegraphics[clip,width=0.9\columnwidth]{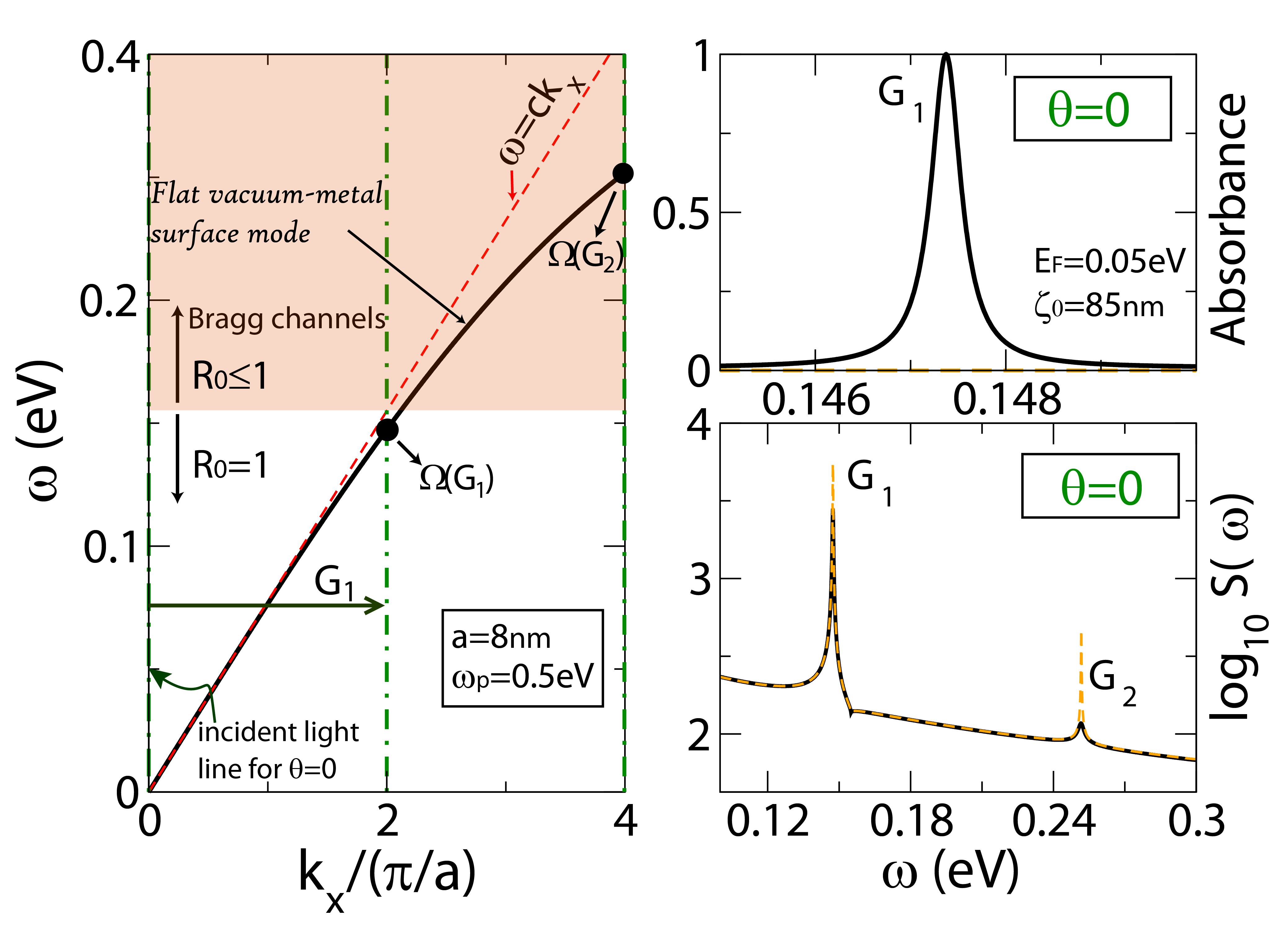}
\par\end{centering}

\caption{\label{fig:02}The SPP spectrum for a flat-surface interface $\Sigma$
with $\hbar\omega_{p}=0.5$~eV and $\gamma=0$ (solid line in the
left panel). Its intersection with the displaced incident photon line
($k_{x}\rightarrow k_{x}+G_{n})$ gives an estimate of the frequency
of SPPs in the $n$th band {[}e.g.,~$\hbar\Omega(G_{1})\simeq0.147$~eV{]}.
Absorption at normal incidence near the first SPP resonance (right
top panel), and the scattering function $S(\omega)$ for a wider frequency
interval (right bottom panel). Dashed lines in the right panels refer
to the bare interface (in the absence of graphene).}
\end{figure}

\begin{figure}
\begin{centering}
\includegraphics[clip,width=0.9\columnwidth]{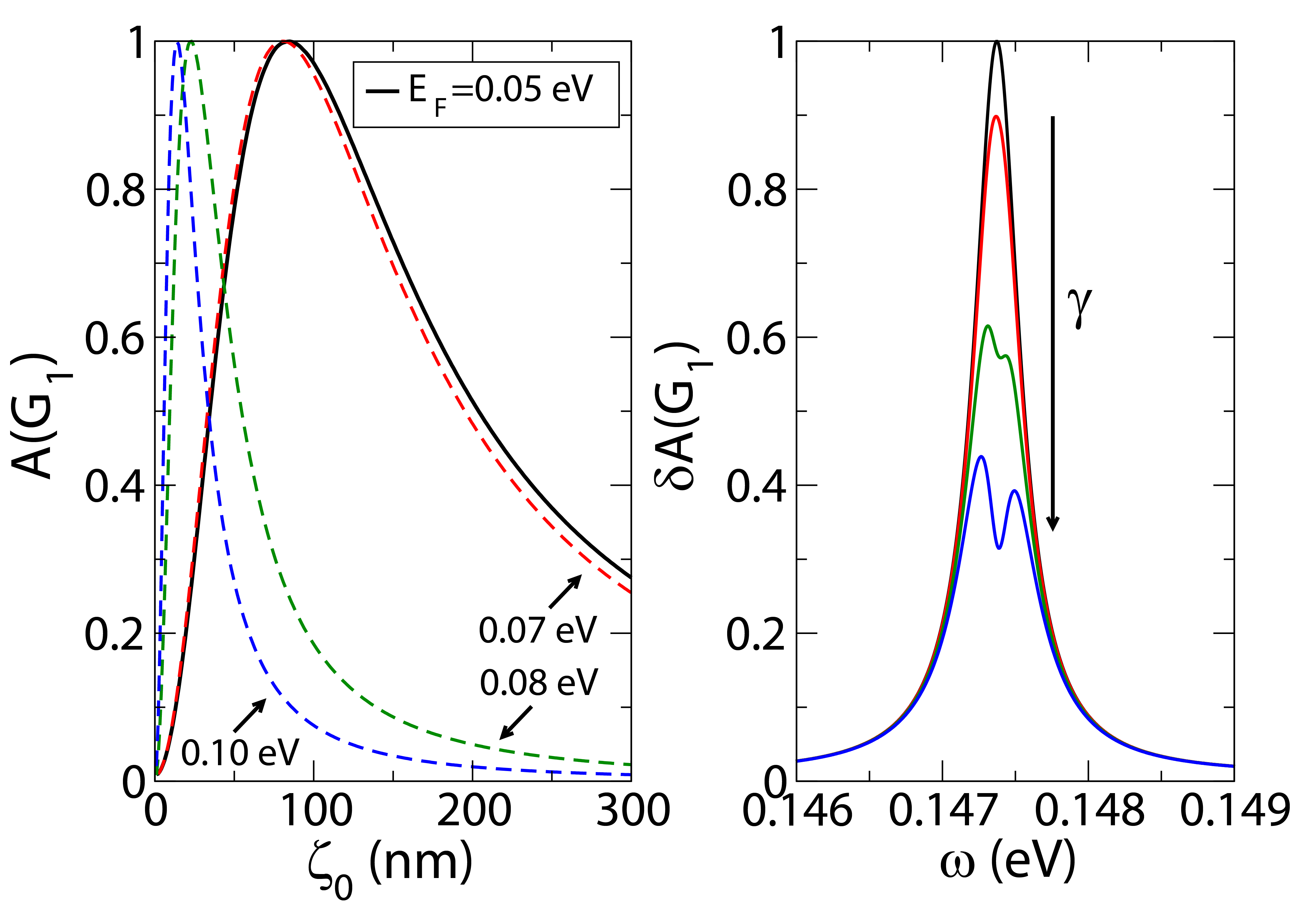}
\par\end{centering}

\caption{\label{fig:03}Dependence of the absorption at resonance on the corrugation
height (left panel) for different graphene Fermi energies, $E_{F}$,
and relative absorption for a nonzero metamaterial relaxation's rate
$\gamma$ (right panel) with $\gamma/\omega_{p}=\{10^{-4},5\cdot10^{-4},10^{-3}\}$
(magnitude increasing as indicated by the arrow). (Left panel) The
fast change in $A(G_{1})$ as $E_{F}$ varies from $0.07$ to $0.08$~eV
results from a suppression of interband transitions in graphene as
$2E_{F}\gtrsim\hbar\Omega(G_{1})$. (Right panel) Appearnce of two
peaks for the highest values of $\gamma$ witnesses a considerable
shift of the resonant position of $A(G_{1})$ relative to $A_{\textrm{bare}}(G_{1})$.
Other parameters as in Fig.~\ref{fig:02}.}
\end{figure}

Coating the metamaterial interface with graphene introduces new channels
for SPP dissipation, as encoded in the optical conductivity $\sigma(\omega)$,
due to Drude diffusive scattering and interband transitions.\cite{colloquium}
Based on general arguments,\cite{Herminghaus1994} we expect that
if the SPP decay rate, $\Gamma_{\textrm{g}}$, introduced by graphene,
matches the broadening due to radiative decay, $\Gamma_{\textrm{rad}}$,
the interference between the specular channel and the high-order Bragg
beams suppresses the reflectivity, hence allowing the SPPs to be fully
relaxed within the graphene sheet. The latter is borne out in Fig.~\ref{fig:02}
(right top panel), which shows the absorption near the lowest SPP
resonance for a sinusoidal corrugation with $\zeta_{0}=85$~nm and
$a=8$~$\mu$m, and at fixed Fermi energy, $E_{F}=50$~meV. We note
that by varying the electronic density of the graphene sheet, it is
possible to obtain complete absorption for $\zeta_{0}/a$ as low as
$10^{-3}$ (see Fig.~\ref{fig:03}). We thus predict that small corrugation
amplitudes are needed for obtaining complete light absorption.

\section{Discussion\label{sec:Discussion}}

In order to understand the reported SPP-assisted enhanced absorption,
we first characterize the spectrum of SPPs in the hybrid graphene-metamaterial
system. The effect of a periodic spatial profile can be qualitatively
appreciated by analogy with one-dimensional electrons subjected to
a weak periodic potential: a length scale is introduced ($a)$, and
the SPP spectrum acquires a multiband structure with period $2\pi/a$,
with gaps opening at the Brillouin zone boundaries $\pm\pi n/a$.\cite{Yuli2012}
Indeed, SPP modes can be excited by photons via coupling through lattice
reciprocal vectors, $G_{n}=2\pi n/a$, with $|n|>0$, whenever the
frequency of the incident photon matches the frequency $\Omega$ of
a SPP in a given band, that is, $\omega=\Omega(q_{n})$. For shallow
spatial profiles, the SPPs frequencies can be estimated by folding
the flat-surface spectrum in the absence of graphene,\cite{comment_spectrum_withgraphene}
$k=(\Omega/c)\sqrt{\epsilon(\Omega)/[\epsilon(\Omega)+1]}$ {[}with
$\epsilon(\Omega)<-1${]}, into the Brillouin zone and locating the
intersection with the incident photon line\cite{Economou1969,Agranovich1982,Weber_Mills_1983,Zayatsa2005}
(or equivalently, by the procedure illustrated in the left panel of
Fig.~\ref{fig:02}). For a rigorous determination of the SPP spectrum
for arbitrary corrugation profiles, we examine the poles of the scattering
matrix $\hat{S}$, defined by $(\boldsymbol{\mathcal{H}}^{+},\boldsymbol{\mathcal{H}}^{-})^{T}=\hat{S}(\boldsymbol{\mathcal{H}}^{\textrm{in}},0)^{T}$,
where $\boldsymbol{\mathcal{H}}^{\pm}=(\mathcal{H}_{-N+1}^{\pm},...,\mathcal{H}_{N}^{\pm})$
and $\boldsymbol{\mathcal{H}}^{\textrm{in}}$ are 2$N$-dimensional
vectors containing the incoming (outgoing) modes in the vacuum (substrate),
and the input field amplitudes, $\mathcal{H}_{n}^{\textrm{in}}=H_{\textrm{in}}\delta_{n,0}$,
respectively. The SPPs frequencies for a given wavevector $k_{x}$
can be obtained by localizing the peaks of the function $S(\omega)=\sum_{n,m}|S_{nm}|$.\cite{WCTan1999}
The scattering function for normal incidence ($k_{x}=0$) is given
in the right bottom panel of Fig.~\ref{fig:02}. Clearly, the SPPs
frequencies are not significantly altered by the presence of graphene
(as seen by the agreement of the peaks positions in both the solid
and dashed lines\cite{comment_spectrum_withgraphene}). Moreover,
these frequencies agree fairly well with the flat-spectrum estimate,
shown in the left panel, based on perturbation theory arguments. The
effect of graphene is therefore to change the spectral weight of scattering
function $S(\omega)$ around the vacuum-metal resonances, $\omega\simeq\Omega(G_{n})$,
without affecting too much their position (note that in Fig.~\ref{fig:02}
only the first two resonances, $|n|=1,2$, are shown). The little
sensitiveness of the resonant frequency of SPPs to the graphene sheet
characteristics turns out to be of significance for experimental realizations.
We can imagine that by changing $\epsilon(\omega)$, for example,
by considering a metamaterial with an appropriate plasma frequency,
the resonant frequencies $\Omega(G_{n})$ can be controlled within
a broad range of values, thus tailoring the hybrid structure for a
desired spectral region. Another possibility is to probe the Fermi
energy of graphene by optical measurements, either by monitoring changes
in the reflectivity or by measuring small changes in the SPPs frequencies. 

\begin{figure}
\begin{centering}
\includegraphics[width=0.8\columnwidth]{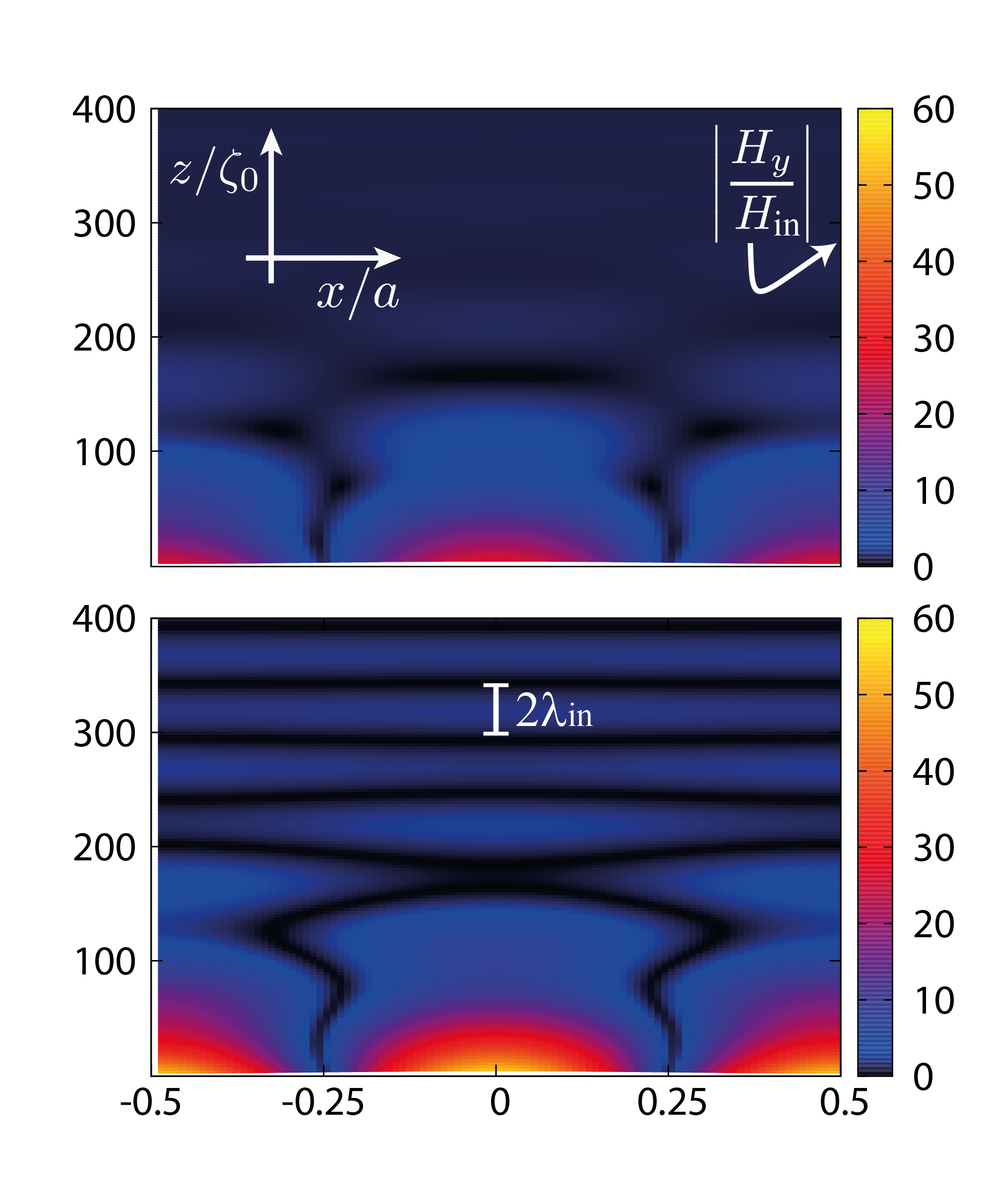}
\par\end{centering}

\caption{\label{fig:04}Contour plot for the magnetic field modulus in the
hybrid graphene-metal corrugated interface (top) and for the same
structure in the absence of graphene (bottom). The field modulus is
periodic in $x$ and thus only the restricted interval $|x|\le a/2$
is shown. The corrugation height is $\zeta_{0}=85$~nm and the impinging
light has $\hbar\omega=0.1474$~eV. The bar indicates the scale of
the oscillation resulting from interference of the incident field
and the reflected field, $2\lambda_{\textrm{in}}=4\pi c/\omega$.}
\end{figure}

The dependence of the absorption line-shape on the corrugation amplitude
shown in Fig.~\ref{fig:03} (left panel) is similar to that found
in metallic nanopatterned gratings with considerable ohmic losses.\cite{Weber_Mills_1983,WCTan1999}
Initially, as $\zeta_{0}$ increases, so does the coupling of SPPs
to the radiation continuum, hence originating a monotonous increase
in the absorption at resonance, $A(G_{1})$. At some point, the radiative
decay balances the SPP damping in the vicinity of graphene, $\Gamma_{\textrm{rad}}=\Gamma_{\textrm{g}}$,
and the reflectivity is completely suppressed (absorption is maximum).\cite{WCTan1999,Herminghaus1994}
Increasing $\zeta_{0}$ beyond such value overcouples the SPP to the
radiation field, and the specular wave is no longer canceled out by
the reradiated field: $A(G_{1})$ decreases. Eventually, $\Gamma_{\textrm{rad}}\gg\Gamma_{\textrm{g}}$,
and absorption becomes negligible, with all the energy being carried
by the reflected wave. Evidently, the value of $\zeta_{0}$ leading
to complete light absorption depends on the Fermi energy of graphene,
since the latter determines the effective value of $\Gamma_{\textrm{g}}$.
For instance, for $E_{F}=0.1$~eV, the maximum absorption occurs
for very small $\zeta_{0}$ (see Fig.~\ref{fig:03}), and given that,
at lowest order, $\Gamma_{\textrm{rad}}\sim\zeta_{0}^{2}$ (see, e.g.,~Ref.~\onlinecite{Mills_Weber_1982}),
we can infer that the SPP damping in graphene $\Gamma_{\textrm{g}}$
is particularly small in such a case. 

The calculation of the field intensity distribution in the vicinity
of the surface $\Sigma$ further confirms the role played by the SPP
decay channel introduced by graphene: In Fig.~\ref{fig:04}, we plot
the magnetic field intensity in the vacuum for incident light in resonance
with SPPs in the first band, $|n|=1$. When graphene is absent (bottom
panel), the large field enhancement effect typical of SPPs in corrugated
interfaces is clearly observed,\cite{Mills_Weber_1982,Weber_Mills_1983}
with $|H_{y}|\approx60|H_{\textrm{in}}|$ in the near field. Here,
the reflectivity is total as inferred by the sharp field interference
pattern observed for $z\gg\zeta_{0}$. When graphene is introduced
(top panel), the far-field interference pattern disappears, demonstrating
the suppression of the reflected wave, while the near field displays
more modest field enhancements, due to SPP damping in the graphene
sheet, in consistency with arguments given previously.

\section{Final Remarks\label{sec:Final_Remarks}}

We finish our discussion with a brief analysis of how losses in the
metallic substrate\emph{ }and modulation of the graphene conductivity
may affect the optical properties of the graphene-metamaterial hybrid
system\emph{. }

\emph{Losses in the metamaterial}. When the dieletric function of
the substrate $\epsilon(\omega)$ is assumed to be a real function,
absorption in the graphene sheet reaches 100\% efficiency when adequate
choices of the corrugation height are made (refer to the left panel
in Fig.~\ref{fig:03}). On the other hand, in any realistic scenario,
a trade-off between absorption in the graphene sheet and damping within
the engineered substrate is expected to happen. In order to quantify
the impact of a non-zero damping parameter $\gamma$ (and hence a
complex dieletric function) on the absorption efficiency of graphene
paired up with the metallic substrate, we compute the change in the
absorption, $\delta A=A-A_{\textrm{bare}}$. Here, $A_{\textrm{bare}}$
denotes the absorbance of the bare metamaterial interface (i.e.,~without
graphene). (Note that subtracting $A_{\textrm{bare}}$ to the absorbance
of the hybrid system is necessary in order to correctly infer the
amount of energy absorbed by the graphene sheet.) The right of Fig.~\ref{fig:03}
shows $\delta A$ for a specific configuration ($E_{F}=0.05$~eV
and $\zeta_{0}=85$~nm) and a few values of the damping rate $\gamma$.
In these cases, and also more generally, we have found that damping
frequencies as low as $\gamma\lesssim10^{-3}\omega_{p}$ are required
so that the graphene sheet is able to absorb considerable amounts
of light ($\delta A\gtrsim0.1$). With this respect, we recall that
in the infrared range, ratios $\gamma/\omega_{p}$ of the order of
$10^{-3}$ are achievable in metals as copper,\cite{Ordal1985} and
hence the restriction on the substrate losses should not impose any
fundamental limitation on the plasmonic performance of the hybrid
system. 

\emph{Modulation of the conductivity}. When graphene adheres to a
corrugated surface its local optical response becomes sensitive to
local perturbations (e.g., via position-dependent strain). However,
in the present case, the SPP dispersion relation of the hybrid system
is practically unaffected by graphene's intrinsic conductivity,\cite{comment_spectrum_withgraphene}
and thus the impact of modulations in $\sigma(\omega)$ in the optical
properties of the hybrid system is expected to be weak. In order to
assess this assumption, we have considered a conductivity profile
$\sigma(x,\omega)$, with the periodicity of the corrugated surface
and magnitude variations of the order of 100\%, namely, $\sigma(x,\omega)=\sigma(\omega)[1+\cos(2\pi x/a)]$,
and no qualitative changes relative to the unmodulated case have been
found; for instance, the peak of the absorbance curve $A=A(\omega)$
in Fig.~\ref{fig:02} is seen to decrease by only about 0.1\% and
shifted by even a lesser amount.

\section{Conclusion}

We have shown that complete light absorption occurs in continuous
sheets of graphene placed on top of simple plasmonic nano-structures.
The physical mechanism on the basis of the two orders of magnitude
boost in the graphene's capability to absorb light demonstrated here
lies in the efficient excitation of SPPs in the periodically corrugated
interface of the hybrid graphene-metamaterial system. Absorbances
far exceeding the graphene's bare value (of about $2\%$)\cite{Nair2008}
are still achievable when intrinsic losses in the metamaterial are
taken into account, provided that the damping rate is much smaller
than relevant frequency scales, namely, the frequency of light and
the plasma frequency of the metamaterial. In this so-called relaxation
regime, $\gamma\ll\omega\ll\omega_{p}$, the plasmonic resonances
of the hybrid system are determined by the plasma frequency of the
metamaterial, allowing the frequency range for enhanced absorption
to be set without the need of engineering graphene. Furthermore, the
plasmonic-enhanced absorbance effect is shown to occur in a wide range
of electronic densities of the graphene sheet (either low- or high-
doped samples). The properties of the hybrid graphene-metamaterial
system described in this work open interesting possibilities, such
as the capture of light and further guidance and control of the excited
plasmonic waves (e.g.,~by designing SPP propagation paths by inhomogeneous
doping of the graphene sheet\cite{Ashkan}), making this system a
potential candidate for plasmonics applications, such as light-harvesting
and photonic circuit implementation.

\section{Acknowledgements}

This work was supported by the National Research Foundation\textendash{}Competitive
Research Programme award \textquotedblleft{}Novel 2D materials with
tailored properties: beyond graphene (Grant No. R-144-000-295-281).
A.F. acknowledges stimulating discussions with J. Viana-Gomes during
his visit to the Graphene Research Centre (GRC) and technical assistance
from M. D. Costa (GRC).


\begin{thebibliography}{10}
\bibitem{comment_boundary_condition}In this work, we assume graphene
to be in contact with the metamaterial substrate, in which case there
is only one interface for the electromagnetic problem.
\bibitem{comment_approximation}In deriving Eqs.~(\ref{eq:reflected_transmitted_fieldds})
and (\ref{eq:system_eq}) we have multiplied the surface current term
$J_{\Sigma}$ by the function $\sqrt{1+[\zeta^{\prime}(x)]^{2}}$.
This approximation is extremely accurate for shallow corrugations
($\zeta_{0}/a\ll1$), and permits to obtain simpler expressions. For
the parameters considered in this work, the maximum absolute error
in the absorption line-shape by adopting this procedure is of the
order of $10^{-3}$.
\bibitem{comment_convergence}For the sinusoidal profile, the $\Upsilon$
functions have the explicit solution: $\Upsilon_{m}^{\alpha\beta}(p)=I_{p}\left[-\alpha\beta\kappa_{\alpha}(q_{m})\right]$,
where $I_{p}(x)$ denotes the Bessel function of the second kind.
Taking $N=6$ is sufficient to obtain accuracy up to 6 figures for
large corrugation amplitudes, $\zeta_{0}/a\sim0.1$. The accuracy
is even larger (up to 12 figures) for the amplitudes considered in
this work.
\bibitem{comment_spectrum_withgraphene}The explicit calculation shows
that the SPP dispersion relation of a flat vacuum-metal interface
is not altered significantly by the inclusion of a graphene sheet
at the interface whenever the condition $(|\sigma_{xx}|/\varepsilon_{0}c)|\left[\epsilon(\omega)+1\right]^{-1/2}|\ll1$
is met. Since the prefactor $|\sigma_{xx}|/\varepsilon_{0}c$ is of
the order of the fine-structure constant, the latter condition is
satisfied for $|\epsilon(\omega)|\gg1$ (that is, $\omega\ll\omega_{p}$).
\end{thebibliography}
\end{document}